\documentclass[journal, letterpaper]{IEEEtran}
\usepackage{cite}
\usepackage{amsmath,amssymb,amsfonts}
\usepackage{algorithmic}
\usepackage{graphicx}
\usepackage{textcomp}
\usepackage{hyperref}
\usepackage{soul}
\usepackage{caption}

\usepackage{url}

\usepackage{textgreek}	
\usepackage{listings}
\usepackage{csvsimple}
\usepackage{longtable}

\begin{document}

	\title{Using network structure and community detection to discover important website features when distinguishing between phishing and legitimate ones}
	\author{Arash Negahdari Kia, Finbarr Murphy, Zahra Dehghani Mohammadabadi, \& Parisa Shamsi}

	\maketitle

\begin{abstract}
	In this paper, we uncover the essential features of websites that allow intelligent models to distinguish between phishing and legitimate sites. Phishing websites are those that are made with a similar user interface and a near similar address to trustworthy websites in order to persuade users to input their private data for potential future misuse by attackers.
    Detecting phishing websites with intelligent systems is an important goal to protect users, companies, and other online services that use the HTTP protocol. An intelligent model needs to distinguish features that are important as input to predict phishing sites. 
    In this research, using correlation-based networks, we provide a novel network-based method to find features that are more important in phishing detection.
    The networks are trained and tested on an established phishing dataset. Three different networks are made by partitioning the dataset by its data instance labels. The important features are found by discovering the hubs of these networks, and the results are presented and analysed.
    This is the first time using a network-based approach for feature selection which is a fast and accurate way to do so.
\end{abstract}


\section{Introduction}
	\label{sec:introduction}
    \PARstart{T}{he} internet has become ubiquitous for all private and commercial activity. In many instances, websites may require personal information such as usernames and passwords; and there is a general degree of implicit trust associated with this information transfer. This allows malicious hackers to steal private data for illicit gain. To do so, they can try to trick people by making them think that they are passing their information to a trustworthy website by displaying a fake website with similar characteristics to the legitimate website and a near similar URL address. Such websites are called phishing websites. Many people fall into this trap and face substantial negative consequences \cite{wu2006fighting}. According to the IBM threat index 2020, phishing is the most popular cyber-attack which happens globally \cite{ibm2020}.
    
    Significant efforts have been made to detect phishing websites and prevent such fraud. However, there is yet no clear way to distinguish legitimate websites from phishing websites. Therefore, efforts are focused on methods that can detect phishing websites with higher accuracy.
    
    Phishing detection is approached in a variety of ways. Most contemporary approaches for detecting phishing websites are based on machine learning and intelligent models, like using a classification method on website features. One way to optimize the results of these approaches is to find the most salient features of a website to identify its legitimacy \cite{patil2019detection, mao2019phishing, chiew2019new, rao2020catchphish}. In another research, the effectiveness of finding features in optimally detecting phishing websites was investigated by two methods: Wrapper-based feature selection and correlation-based feature selection \cite{basnet2012feature}. In correlation-based feature selection, the criterion had been calculated, and important subset of features were selected according to the criterion. Wrapper-based feature selection needs the supervised algorithms and labels for each instance of the dataset for selecting the important features. In the wrapper method, a subset of features that make the most accurate prediction/classification is selected. The researchers finally compared the performance of both methods in their study.
    In this paper, we propose a heuristic method to find the most important features of a website to help intelligent models in phishing detection. Knowledge graph representation has helped us find the most important features in distinguishing between phishing and legitimate websites. In our approach nodes of the network represent the features and those nodes that have more connections have more influence on other nodes, and therefore, they represent more important features. In section~\ref{sec:related}, we give a brief explanation and examples of some concepts used in our proposed method.
    
    In section~\ref{sec:pre}, we investigate some preliminaries methods. We explain our proposed method in section~\ref{sec:method} and discuss its results in section~\ref{sec:result}. Section~\ref{sec:conclusion} concludes the paper and presents suggestions for future researches.

    \subsection{Related Works}
    \label{sec:related}
    
    \subsubsection{Phishing Detection}    
        Some considerable research has been undertaken to increase the detection of phishing websites. The approaches can be classified into the blacklist approach and the heuristic approach.
        
        \begin{enumerate}
            \item Blacklist Approach:
            
            In the blacklist approach, a list of malicious URLs is formed as a blacklist. When a user requests a website, the domain will be compared with the list to find a match. If this match is found, the connection would not be allowed. Its disadvantage is that the blacklist should be updated frequently, and some phishing websites may not be discovered.
            
            Another study proposed a blacklist approach that keeps the blacklist up to date by using search engine results to detect suspicious domains \cite{sharifi2008phishing}. This way, the website's legitimacy can be checked. Another research proposed a system (PhishNet) using an algorithm to find a close match in the blacklist \cite{prakash2010phishnet}. Another study proposed an approach that uses the redirection URLs from phishing websites for completing the blacklist \cite{lee2014poster}.
            
            \item Heuristic Approach:
            
            In the heuristic approach, some techniques like machine learning are used to find phishing websites based on general phishing features. The advantage of this approach is that new phishing websites can be detected.
            
            In one study, researchers propose a heuristic method based on a relative detection based on the website's logo and legal logo \cite{yao2018deep}. Another research proposed a phishing detection technique based on machine learning using an analysis of the URL's features, website host and interpretation of the visual appearance \cite{patil2019detection}. Mao et al. propose a heuristic phishing detection method using machine learning techniques to find the similarity between the website's user interface and a legitimate website's user interface \cite{mao2019phishing}. Chiew et al. propose a heuristic feature-based method that uses machine learning to detect phishing websites, called the Hybrid Ensemble Feature Selection (HEFS) \cite{chiew2019new}. HEFS included two steps: The first by using the Cumulative Distribution Function gradient algorithm found the number of optimal features and in the second step, selected a subset of features by the hybrid framework. HEFS had high performance when using a random forest algorithm.
            Rao et al. propose a method based on URL and using TF-IDF property to detect a phishing website. Also, the dataset that we used in our paper has features based on URLs such as URL$\_$length \cite{rao2020catchphish}. Zhang et al. also propose a method only using URL addresses for phishing detection. Techniques used in the method are bidirectional LSTM, skip-gram, and CNN \cite{zhang2020phishtrim}. Chavan et al. propose a phishing detection method using deep learning technique and feature engineering and reduces features of the dataset from 19 to 10 \cite{chavan2020phishing}. One of the big problems in phishing detection is the lack of phishing data against legitimate data. Shirazi et al. used data augmentation to solve the problem \cite{shirazi2020improved}.
        \end{enumerate}
        
        \subsubsection{Network Structures}
        \label{sec:network}
        
            Many real-world phenomena can be modelled by networks such as social networks and information networks.
                
                Social networks show the interaction of people or groups of people in the form of nodes and edges connecting them \cite{newman2010networks}.
                Barabasi et al. discuss scientific collaborations as complex networks \cite{barabasi2016network}. Nekovee et al. propose a model to show the spread of rumours \cite{nekovee2007theory} and Potts et al. propose a market-based definition of creative industries, both based on complex social networks \cite{potts2008social}. In another research, Schimit used complex networks for modelling a population to show how people connect and analysed it as a disease spreading model \cite{schimit2018disease}.
                
                Information networks are networks showing interactions between some items of data. Information networks are a type of network showing the interaction between concepts in the outside world that can be interpreted as nodes and links \cite{newman2010networks}. We can refer to the World Wide Web as the best-known information network.
                A citation network is a network that is based on paper citations. Son et al. propose a method for an academic paper recommender system based on citation networks \cite{son2018academic}. Their proposed network is a multilevel simultaneous citation network, and this method is useful when citation information is not enough.
                
        
            In this paper, we analyse the information network built from the phishing and legitimate websites data for feature selection. The features can be used for better phishing detection models used in both literature and application like intrusion detection systems.

        \subsubsection{Community Detection}
        
            Community detection is a procedure used to group network nodes in a way to make nodes in each community have dense connections. As a result, a better understanding of the network's structure and function is discovered. There is a broad application of community detection used in the researches. Kanavos et al. propose an efficient methodology for community detection to analyse the behaviour of users on an emotional level based on their tweets \cite{kanavos2018emotional}. In this paper, we use community detection to cluster the website features in order to analyse their similarities. We deploy a similarity knowledge graph using different characteristics and features of phishing/legitimate websites. This is a unique approach that has not been used in the phishing detection research area by far. Network modelling has been found useful and effective in different areas of research, and this is the first time it is used in phishing research.
            
    \subsection{Preliminaries}
    \label{sec:pre}
    
        \subsubsection{Constructing a Similarity Graph}
        \label{sec:constructingsimilaritygraph}
            
            We define
            
            \begin{equation}
                Correlation (feature_{i}, feature_{j}) = 1 - \frac{6 \sum d_{k} ^ 2}{n (n ^ 2 - 1)},
                \label{eqn:correlation}
            \end{equation}
            
            \noindent where $d_{i}$ is the difference between two ranks of each feature for each instance of the dataset, and $n$ is the number of instances in the dataset. By using \equationautorefname{ \ref{eqn:correlation}}, we form the correlation matrix. The dataset features are categorical, so we use Spearman's rank-order correlation \cite{kumar2018correlation}.
            
            The distance between $feature_{i}$ and $feature_{j}$ can be obtained from \equationautorefname{ \ref{eqn:d}}.
            
            \begin{equation}
                d_{i, j} = \sqrt{2\ (1 - Correlation(feature_{i}, feature_{j}))}.
                \label{eqn:d}
            \end{equation}
            
            Finally, we use \equationautorefname{ \ref{eqn:similarity}} to form the similarity matrix.
            
            \begin{equation}
                similarity\ measure = e ^ {-d_{i, j}}.
                \label{eqn:similarity}
            \end{equation}
            
            This approach of constructing a similarity matrix is used in other researches such as the researches of Bonanno et al. \cite{bonanno2003topology}, Wang et al. \cite{wang2018correlation}, and Song et al. \cite{song2011evolution}. Most of these researches are in the financial data mining and analysis domain. Our research employs the same models to determine the existence of phishing.
            
        \subsubsection{Louvain Community Detection}
        \label{sec:louvain}
            Community detection determines the similarity amongst features of the phishing dataset.
            Louvain is a greedy and extendable community detection method that divides a large network into communities \cite{blondel2008fast}, and because of its greedy nature, it is a fast method in comparison with other methods, specially when dealing with complex networks  \cite{chejara2017comparative}.
            Louvain is based on optimizing the modularity meaning the detection of communities in a way that nodes in a community have dense connections, while nodes in different communities have scattered connections. The Louvain algorithm is described as following:
            
            \begin{enumerate}
                \item Consider each node as a community.
            
             \item Merge two communities if it raises the modularity.
            
             \item Repeat step 2 until no other changes could be done, and that means the modularity is optimized.
            \end{enumerate}
            
        \subsubsection{Maximum Spanning Tree (MST)}
        \label{sec:mst}
           
           Using the maximum spanning tree helps us find the strongest relationship structure amongst the features of our phishing dataset when modelled into a correlation network.
           A maximum spanning tree ($T(V_{T}, E_{T})$) is a subgraph of an edge-weighted undirected graph ($G(V_{G}, E_{G})$) that,
            
            \begin{equation}
                V_{T} = V_{G},
            \end{equation}
            
            \noindent and
                
            \begin{equation}
                E_{T} \subset E_{G},
            \end{equation}
            
            \noindent with the maximum possible total edge weight where $V_{T}$ and $E_{T}$ are sets of the tree's vertices and edges and $V_{G}$ and $E_{G}$ are sets of the graph's vertices and edges.
        
            We will use Kruskal's algorithm \cite{kruskal1956shortest} to form the maximum spanning tree. The algorithm is described as following:
        
        \begin{enumerate}
            \item Sort the graph's edges in descending order.
            
            \item Pick the first edge.
            
            \item Pick the next edge if the set of selected edges up to this step does not form a cycle. A cycle is a non-empty trail in which the only repeated vertices are the first and last vertices \cite{bender2010lists}.
            
            \item If the number of selected edges is one unit less than the number of the main graph's vertices, stop the algorithm. Else, repeat step 3.
        \end{enumerate}    
            
        Using thresholding instead of a maximum spanning tree may lead to expert bias. Even using statistical significance testing would need distribution assumption which would also lead to expert bias.
            
        \subsubsection{Centrality Measures Used In The Research}
                \label{sec:hub}
            Centrality measures are used to find the most important nodes in a knowledge graph. There are many centrality measures defined in the network science. The meaning of what is important in this context depends on the mathematical definition of each centrality measure \cite{newman2010networks}. 
            These measurements help us capture feature attributes of the phishing dataset. These measurements help us understand which features are more important than others in the phishing dataset. 
            
            \begin{itemize}
                \item Degree:
                
                In a graph, the number of edges connected to each node is called its degree. In an undirected graph (G=(V, E)), the relationship between the number of edges (E) and the number of nodes (V) is
                
                \begin{equation}
                    \sum _{v\in V}\deg(v)=2|E|.
                    \label{eqn:degree sum formula}
                \end{equation}

            \item Hub:

                To find the most important nodes in a graph, we can use different measures and definitions. For example, we can say if a node's degree is higher, it is more influential in the graph. In network science, a node with a degree much higher than the average is called hub \cite{barabasi2016network}.
                
                In this research, we consider nodes with a degree higher than two as a hub.
            \end{itemize}
            
        \subsubsection{Gamma Value}
        \label{sec:gamma}
        
            Gamma value is a measurement in network structures that shows the scale-freeness of the network. 
            In some networks, connections between nodes are based on a power-law distribution called preferential attachment. In these networks, called scale-free networks, the gamma value in \equationautorefname{ \ref{eqn:gamma}} is a parameter in the range $2 < \gamma < 3$ \cite{newman2010networks}. Social networks are a kind of scale-free networks. In social networks, there are few nodes with dense connections and many nodes with few connections.
            In the case of higher gamma values, there will be fewer hub nodes with higher degree and more nodes connected to the hubs with less degrees. This means that in higher gamma values, we have some important features and many other features that relate to these hub features. Therefore, it may be possible that they can be ignored when constructing intelligent phishing detection models.
            Imagine a network with $n$ nodes. If the number of nodes with degree k is $n_{k}$, then the probability that a node is of degree k is equal to
            
            \begin{equation}
                p(k) = \frac{n_{k}}{n}.
            \end{equation}
            
            The proper distribution function for the above expression in a network is as follows:
            
            \begin{equation}
                P(k) \sim k ^ {-\gamma}.
                \label{eqn:gamma}
            \end{equation}
            
            In this paper, first, we will calculate the gamma value for each network structure constructed from the phishing dataset. Subsequently, we provide a network analysis of the nodes and their connections to discover important nodes which correspond to important features of websites.
    	
 \section{Method}
    \label{sec:method}
        
\begin{figure}[htbp]
            \centering
            \includegraphics[width=0.4\textwidth]{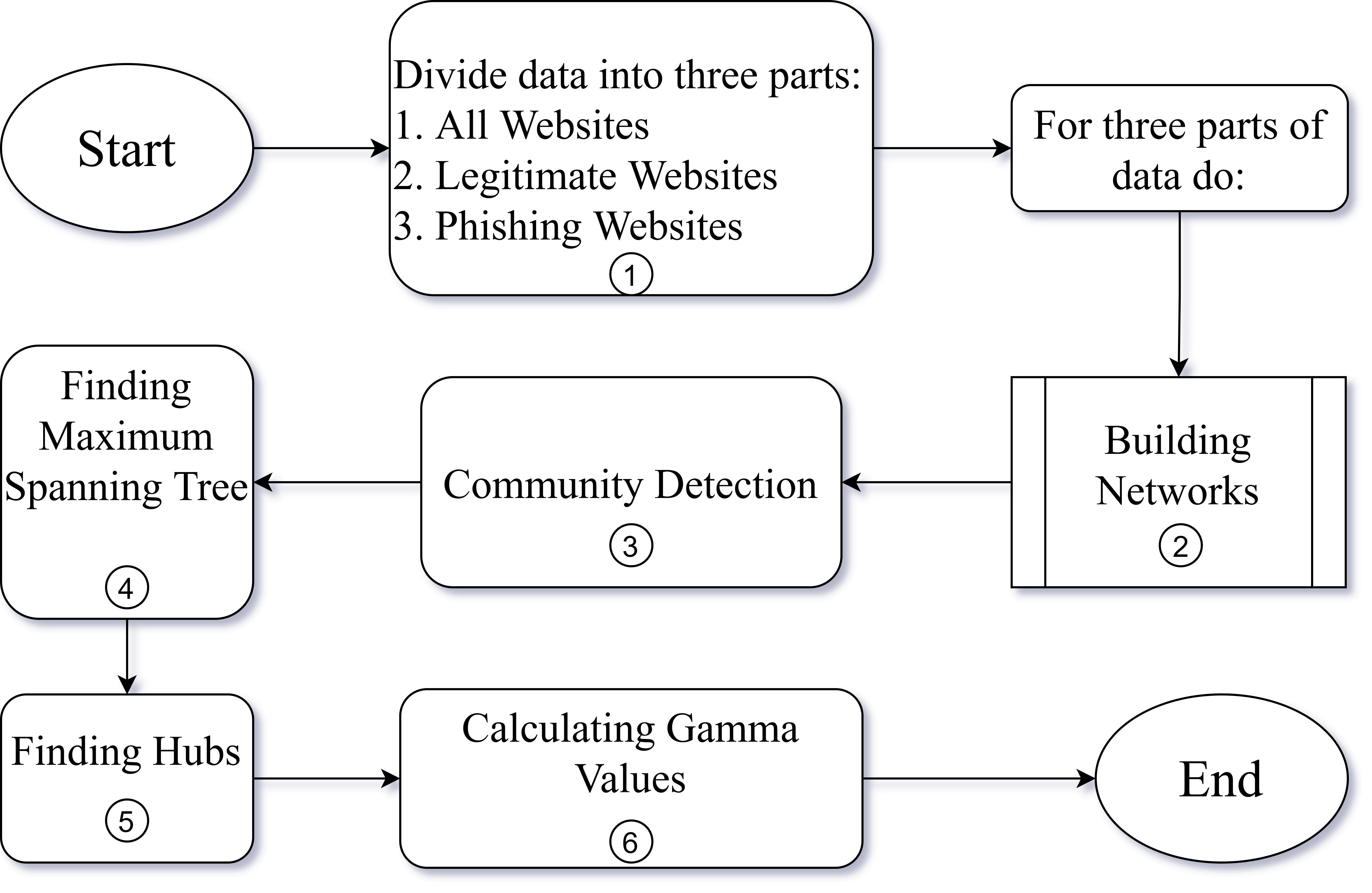}
            \caption{An overview of the research methodology}
            \captionsetup{justification=centering}
            \label{fig:researchflowchart1}
        \end{figure}

        In this section, we explain the design of the study, as shown in \figurename{ \ref{fig:researchflowchart1}} and network construction mechanism in \figurename{ \ref{fig:researchflowchart2}}. We apply a process on the salient features of a website and find the most effective ones that can help intelligent models to detect phishing websites. These features are described in the Appendix.
        
        \subsection{Design of the study}
        
        In the following, we describe the research methodology of \figurename{\ref{fig:researchflowchart1}} for each sub-procedure which is enumerated in the figure.
        
\begin{figure}[htbp]
            \centering
            \includegraphics[width=0.4\textwidth]{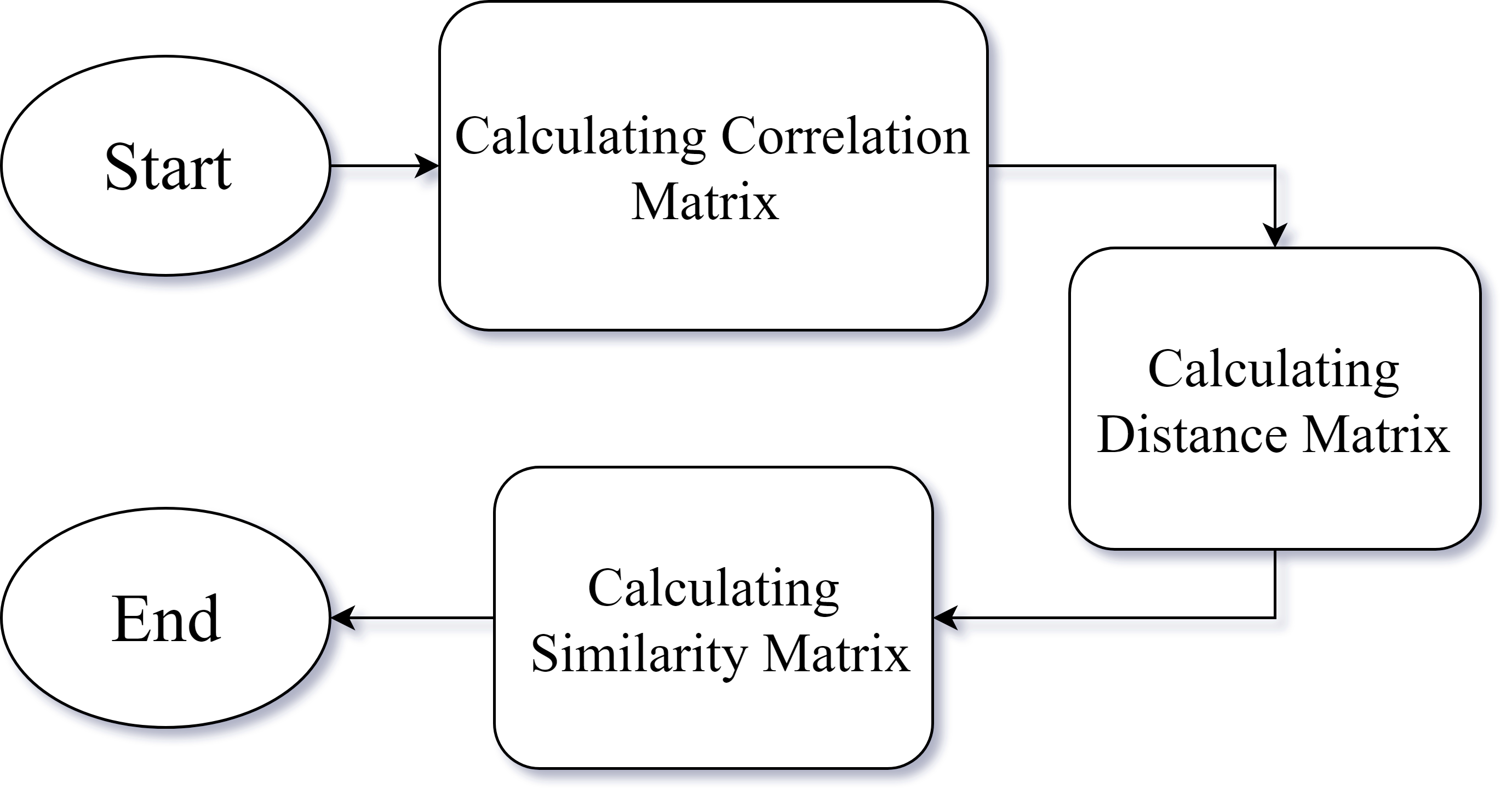}
            \caption{An overview of network building steps in the research methodology}
            \captionsetup{justification=centering}
            \label{fig:researchflowchart2}
        \end{figure}

        \begin{enumerate}
            \item We divided data into three parts. The first group contains all the websites. In the second group, there are only legitimate websites, and the last group includes only phishing websites. 
            All the features are the same in the three parts of the data.
            The following steps apply to all three groups. For each, we build a network from the dataset features. This is to capture the characteristics of different categories of websites along with a network for all the websites together.
            
            \item For building the networks, as it can be seen in \figurename{ \ref{fig:researchflowchart2}}, first, the correlation matrix must be calculated as described in section \ref{sec:constructingsimilaritygraph}. Then, we calculate the distance matrix, and by doing so, we calculate the similarity matrix for each part of the dataset. The reason for doing this is to construct a similarity graph where nodes represent the features of websites in the dataset and the links with their weights represent the similarity between each pair of features. The code for network construction has been added in the GitHub account of the paper\footnote{The code is available here: \href{https://github.com/dmresearches/phishing}{https://github.com/dmresearches/phishing}}.
            
            The network construction procedure has a successful history of representing the similarity between data in finance paradigm. A full and comprehensive description of these correlation networks has been studied in network science literature \cite{bonanno2003topology}.

            \item For finding the features that are most related to each other, we apply a community detection algorithm on all three graphs extracted in the previous step. In this research, we use the Louvain modularity, which is described in section~\ref{sec:louvain}.
            
            \item We find the maximum spanning tree of all three networks built in part 2 of \figurename{\ref{fig:researchflowchart1}}, which the weights of the edges are the similarity between the features (described in section~\ref{sec:mst}). The reason for doing so is to capture the most important feature relationships in each dataset. The maximum spanning tree finds the strongest relations among nodes in each graph.
            
            \item As described in section \ref{sec:hub}, we find hubs for each maximum spanning tree. By doing this, we find the most important features for each category of websites. These features are the most related ones to other features in their dataset. In other words, these hub features can be seen as the candidates in a feature selection procedure for the future supervised or semi-supervised prediction of phishing or legitimate websites. 
            
            \item Finally, we find the gamma values for each network, as described in section~\ref{sec:gamma}. The value shows if the hub features are good representatives of other features. As described in section \ref{sec:gamma}, it is known that in scale-free networks with high gamma values, there are fewer nodes with high degree and a lot of other nodes with low degree. This means that the minority nodes with high degree that we are going to call as hubs are those which relate to many other nodes which represent the features in our dataset.
            
        \end{enumerate}
        
        In the next section, the graphs, trees, and numerical results achieved from our proposed method are presented to discover the essential features in phishing, legitimate, and the whole dataset.
            
    \section{Results and Discussion}
    \label{sec:result}
    
\begin{figure*}[htbp]
            \centering
            \includegraphics[width=1\textwidth]{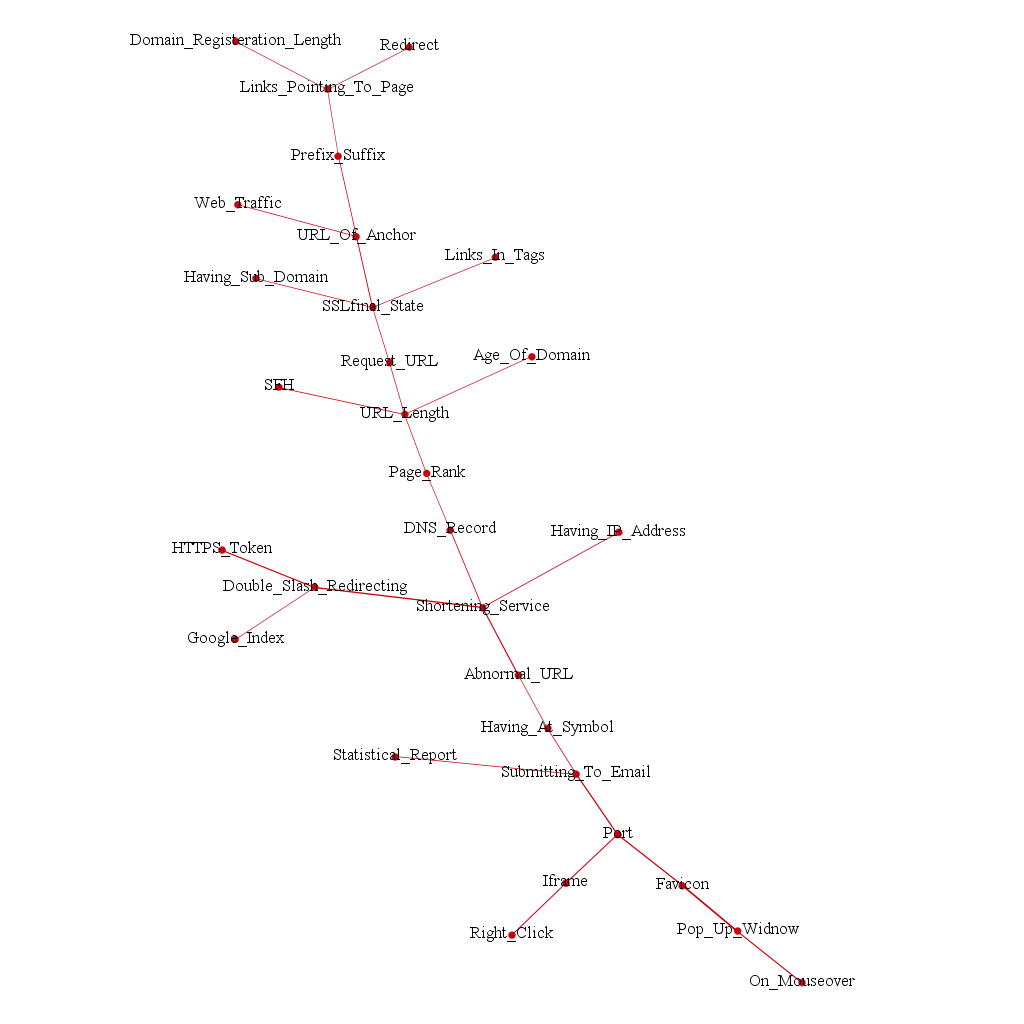}
            \caption{Maximum spanning tree for the graph extracted from the dataset of all websites}
            \label{fig:mstall}
        \end{figure*}

\begin{figure*}[htbp]
            \centering
            \includegraphics[width=1\textwidth]{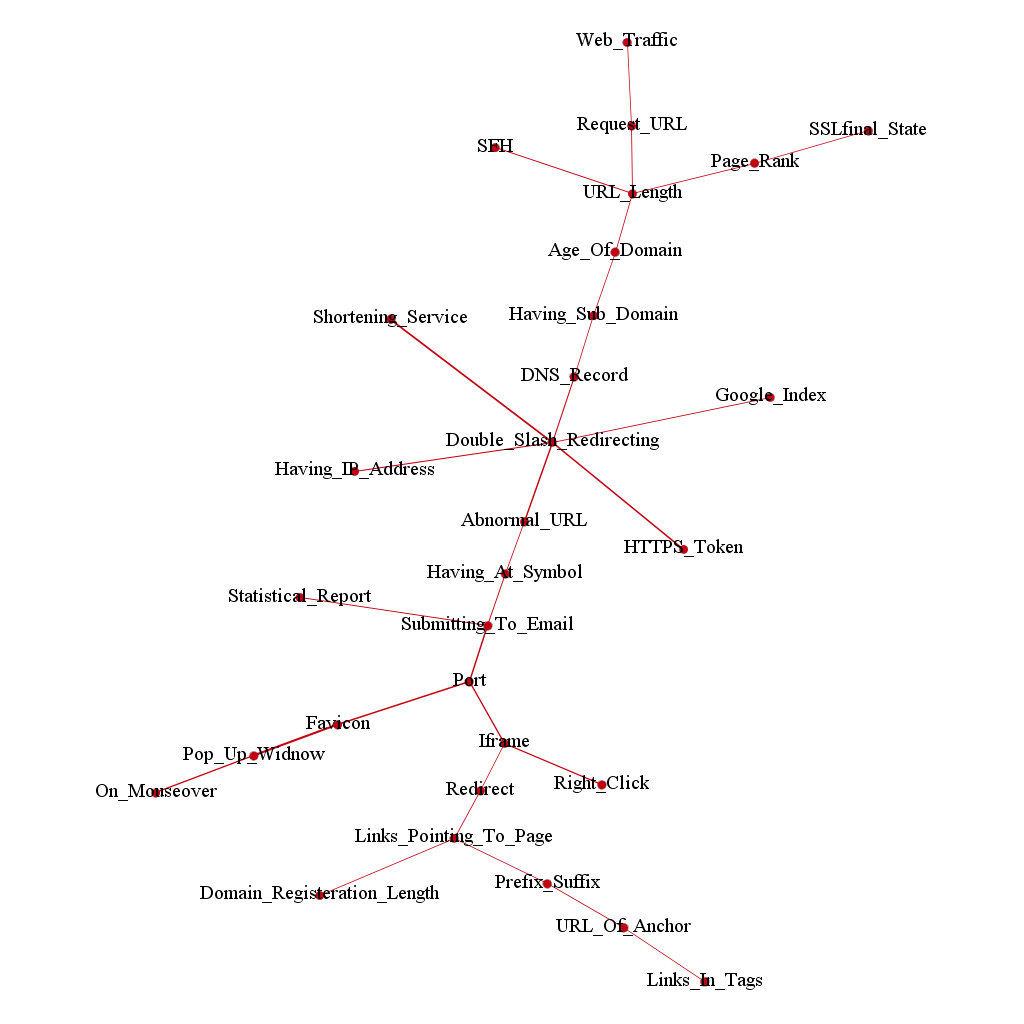}
            \caption{Maximum spanning tree for the graph extracted from the dataset of legitimate websites}
            \label{fig:mstlegitimate}
        \end{figure*}

\begin{figure*}[htbp]
            \centering
            \includegraphics[width=1\textwidth]{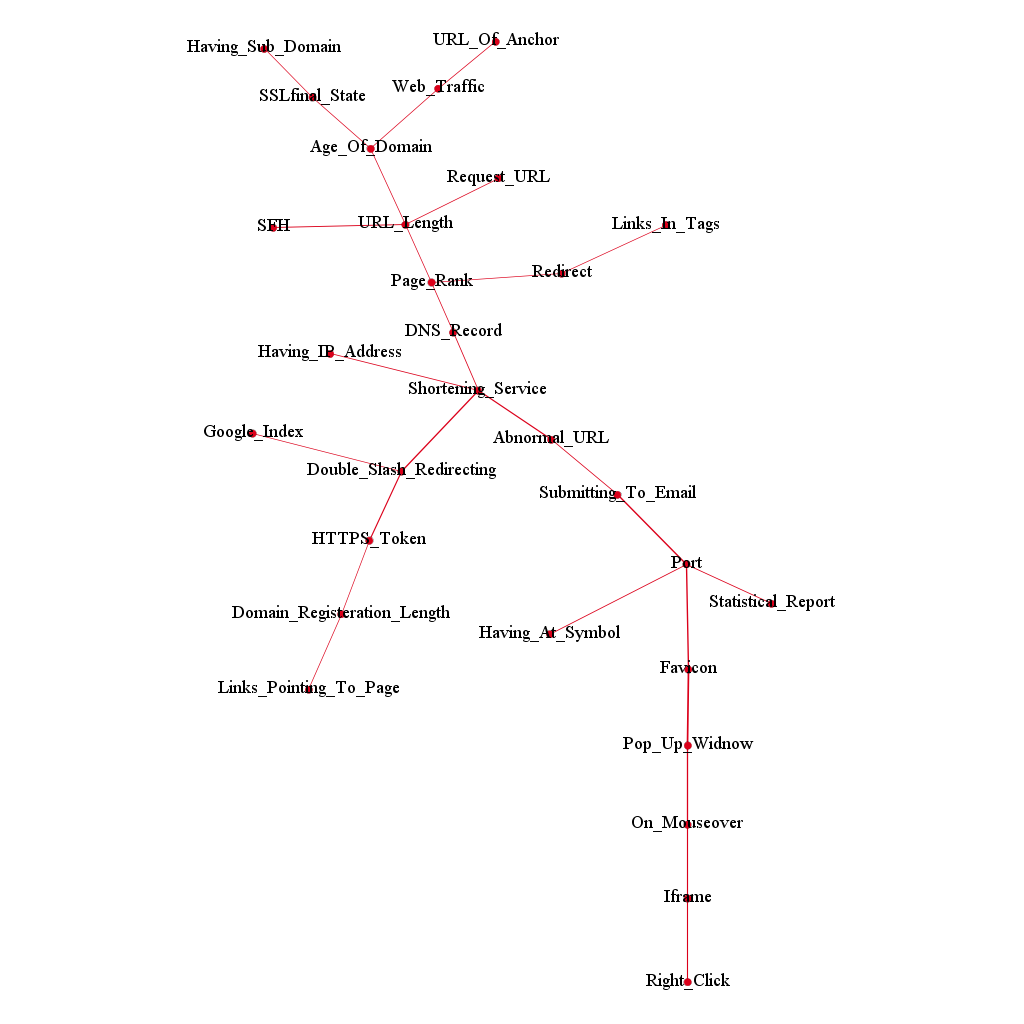}
            \caption{Maximum spanning tree for the graph extracted from the dataset of phishing websites}
            \label{fig:mstphishing}
        \end{figure*}

        In this section, we provide an analysis and discussion based on the results achieved from the methodology outlined in section~\ref{sec:method}. As discussed, we construct a maximum spanning tree for features of each network; all websites presented in \figurename{ \ref{fig:mstall}}, legitimate websites presented in \figurename{ \ref{fig:mstlegitimate}}, and phishing websites presented in \figurename{ \ref{fig:mstphishing}}.
        
        In each maximum spanning tree, we find hub features as described in section~\ref{sec:hub}. These hubs are listed in tables \ref{tab:huball}, \ref{tab:hubligitimate}, \ref{tab:hubphishing}.
            
        \begin{table}[htbp]
            \centering
            \caption{Website features discovered as hubs (nodes with high connections) in the Maximum Spanning Tree built from all the websites}
            \begin{tabular}{|c|c|c|}
                \hline
                \textbf{Hub label}           & \textbf{Degree}      & \textbf{Community} \\ \hline
                    Shortening Service       & 4        & 0               \\ \hline
                    SSLfinal State           & 4        & 2               \\ \hline
                    URL Length               & 4        & 2               \\ \hline
                    Double Slash Redirecting & 3        & 0               \\ \hline
                    Links Pointing To Page   & 3        & 2               \\ \hline
                    Port                     & 3        & 1               \\ \hline
                    Submitting To Email      & 3        & 1               \\ \hline
                    URL Of Anchor            & 3        & 2               \\ \hline
            \end{tabular}
            \label{tab:huball}
        \end{table}
        
        By finding the maximum spanning tree, we are specifying the strongest relations among the features. For example, in \figurename{ \ref{fig:mstall}} which shows the maximum spanning tree of features in the dataset of all websites, if a website has the feature "Shortening Service", it is more probable that it has the feature "Double Slash Redirecting", and so it is more probable that it has the feature "HTTPS Token". In general, it can be said that the features that are hubs are more important than other features of the website in distinguishing between phishing and legitimate, and changes in the values of these features affect the values of other features.
        
        Maximum spanning tree of features in legitimate websites (\figurename{ \ref{fig:mstlegitimate}}) have six hubs listed in table~\ref{tab:hubligitimate}. These features are the most effective features to discuss the legitimacy of a website. In the Appendix we describe in which state of each feature, it is effective in determining the legitimacy of a website.
        
        \begin{table}[htbp]
            \centering
            \caption{Website features discovered as hubs (nodes with high connections) in the Maximum Spanning Tree built from the legitimate websites}
            \begin{tabular}{|c|c|c|}
                \hline
                \textbf{Hub label}          & \textbf{Degree}      & \textbf{Community} \\ \hline
                Double Slash Redirecting    & 6         & 2               \\ \hline
                URL Length                  & 4         & 1               \\ \hline
                Ifram                       & 3         & 0               \\ \hline
                Links Pointing To Page      & 3         & 1               \\ \hline
                Port                        & 3         & 0               \\ \hline
                Submitting To Email         & 3         & 0               \\ \hline
            \end{tabular}
            \label{tab:hubligitimate}
        \end{table}
        
        In \figurename{ \ref{fig:mstphishing}}, the maximum spanning tree of features in phishing websites dataset is displayed whit six features listed in table~\ref{tab:hubphishing} as hubs. That means, for checking if a website is a phishing one, we can focus on these features in prediction models and gain a better performance.

        Some features like "URL Length" are common in both hub features of the legitimate websites and the phishing websites. That means the length of the website's URL, can help us judge both the legitimacy or being a phishing one. This shows that the value of the URL length feature is a discriminator of the classes phishing, or legitimate.

        \begin{table}[htbp]
            \centering
            \caption{Website features discovered as hubs (nodes with high connections) in the Maximum Spanning Tree built from the phishing websites}
            \begin{tabular}{|c|c|c|}
                \hline
                \textbf{Hub label}          & \textbf{Degree}      & \textbf{Community} \\ \hline
                Port                        & 4         & 0               \\ \hline
                Shortening Service          & 4         & 2               \\ \hline
                URL Length                  & 4         & 1               \\ \hline
                Age Of Domain               & 3         & 1               \\ \hline
                Double Slash Redirecting    & 3         & 2               \\ \hline
                Page Rank                   & 3         & 1               \\ \hline
            \end{tabular}
            \label{tab:hubphishing}
        \end{table}

        In table~\ref{tab:gamma}, the gamma values are presented. These gamma values are calculated (as described in section~\ref{sec:gamma}) for features in all websites, legitimate websites, and phishing websites. The higher the gamma value is, the less the number of hubs and the more the degrees of the hubs would be. So, those hubs are better candidates for us to predict the website type (legitimate, or phishing). The gamma value of features in legitimate websites is higher than the gamma value of features in phishing websites. As a result, we can check the legitimacy of a website by the hub features in table~\ref{tab:hubligitimate} with better performance than checking if a website is phishing using the features in table~\ref{tab:hubphishing} because this gamma value is higher and the tree is more scale-free (described in section~\ref{sec:gamma}).

        \begin{table}[htbp]
            \centering
            \caption{Gamma values of the maximum spanning trees for three different datasets of all data, legitimate data, and phishing data. Higher gamma indicates features that make a scale-free network, and the more the hub features are effective in website type prediction.}
            \begin{tabular}{|c|c|}
                \hline
                \textbf{Group}  & \textbf{Gamma values} \\ \hline
                All Of Data     & 0.09                  \\ \hline
                Legitimate Data & 0.13                  \\ \hline
                Phishing Data   & 0.08                  \\ \hline
            \end{tabular}
            \label{tab:gamma}
        \end{table}

        Table~\ref{tab:huball}, table~\ref{tab:hubligitimate}, and table~\ref{tab:hubphishing}, also show some of the results of the community detection algorithm on the networks built in the second part of the method.
        The nature of community detection algorithms is to cluster similar entities into the same cluster/community. Given that, all those features that appear in the same community can be considered having the same importance in phishing/legitimate detection systems so that one can be a delegate for others.
        As you can see in table~\ref{tab:hubligitimate}, most features belong to community 0, so features in this community play a more effective role in detecting legitimate websites. In table~\ref{tab:hubphishing}, most features belong to community 1, so features in this community play a more effective role in detecting phishing websites.
        
        The above method was also applied on different random subsets of the dataset and the results stayed the same, which shows the reliability of the results.

        For evaluating the proposed method, we follow these steps:
        Table~\ref{tab:huball} shows three nodes with the degree of 4. 
        These are the most important features when discussing the legitimacy or illegitimacy of a website. So these are some of the features we worked with. For other nodes with the degree of 3, we chose the ones directly connected to a node with the degree of 4, as can be seen in \figurename{ \ref{fig:mstall}}. As the result, we were dealing with five features: "SSLfinal$\_$State", "Shortening$\_$Service", "URL$\_$Length", "URL$\_$Of$\_$Anchor", and "Double$\_$Slash$\_$Redirecting". We chose the eXtreme Gradient Boosting (XGBoost) \cite{chen2016story} algorithm for classification by the selected features since it is one of the strongest ensemble methods. The accuracy of classification by these five selected features was 0.917. For comparison, we used the Principal Component Analysis (PCA) \cite{pearson1901liii} method by five components for feature selection and ran the XGBoost algorithm for the dataset with the five components. The accuracy by using the PCA method was 0.899.

        When talking about cybersecurity, we are facing a complex system that its elements do not have a linear relationship with each other, and also some of the relationships are not clear. So the best way to model such a system is using network-based approaches. The network-based approaches can discover hidden patterns in the system. Using methods like Principal Component Analysis (PCA) for feature selection and finding the most important features in phishing detection would implement the assumption that the features have a linear relationship with each other, which here is not the case.
        
    \section{Conclusion}
    \label{sec:conclusion}
    
        In order to find the most important features for intelligent models to help them detect phishing websites, we propose a method that finds these features and discovers the connections between them. In this way, we can prevent data loss from the phishing website and provide information security for those who use the HTTP protocol and machines like smart routers that can filter the malicious HTTP traffic.
        
        In this paper, we built correlation-based networks of features in phishing, legitimate, and all websites in our dataset. We subsequently identify important features in each network by finding the hubs in them that had the most effect on the other features.
        
        By extracting the relation networks out of the datasets and finding the hub nodes and gamma values for scale-freeness, we showed which features have a stronger effect on the website class (phishing or legitimate) and which website class is more dependent on particular features. 
        
        In the network made by phishing instances, the important features were, Port, Shortening Service, URL Length, Age Of Domain, Double Slash Redirecting, and PageRank.
        In the network made by legitimate instances, the important features were, Double Slash Redirecting, URL Length, Ifram, Links Pointing to Page, Port, and Submitting To email.
        In the knowledge graph made by the whole dataset, the important features were, Shortening Service, SSLfinal State, URL Length, Double Slash Redirecting, Links Pointing To Page, Port, Submitting TO email, and URL Of Anchor.
        
        The results of our study can be used in smart routers and intrusion detection systems that try to monitor the HTTP traffic and filter the phishing websites.
        
        Future researches that analyses different supervised models with our reduced feature sets for phishing detection using different similarity functions other than correlation can be employed to produce different networks that capture different information from the dataset.

\appendices

\section{Data Gathering}

    The studying dataset is mainly gathered from archives of "PhishingTank", "MillerSmiles" and Google's searching operators \cite{mohammad2016uci} and includes 11055 samples, 30 features, and labels that show, according to the listed features, if a website is phishing or not. The rows including the listed features are categorized as 1, 0, -1. The value 1 defines that a website is legitimate, 0 defines the situation that a website is suspicious to be phishing, and -1 defines that a website is phishing.
    
\begin{figure*}[htbp]
		\centering
		\includegraphics[width=1\textwidth]{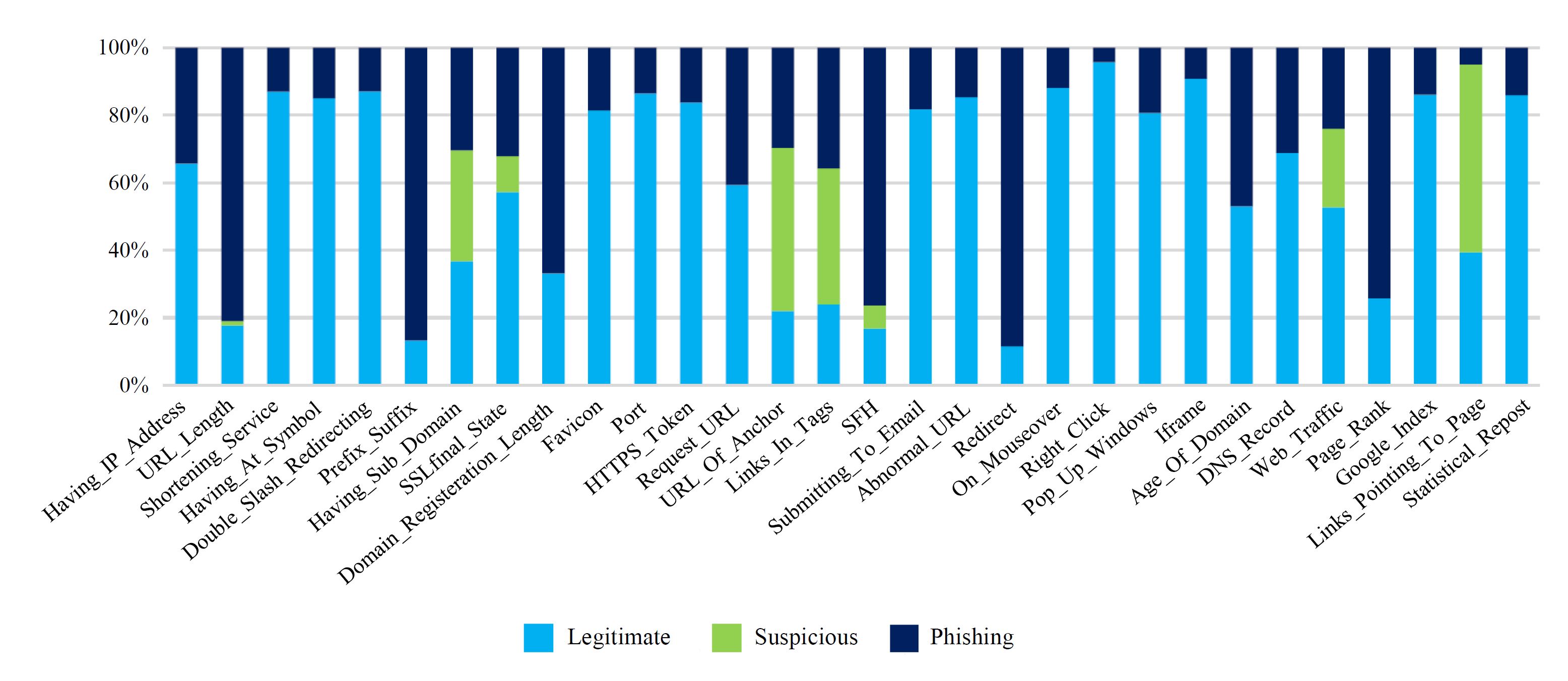}
		\caption{Features in the dataset and the percentage of them appearing in phishing, legitimate, and suspicious classes.}
		\label{fig:features}
	\end{figure*}

	In \figurename{ \ref{fig:features}}, all dataset features can be seen. Also, the percentage that shows, in each feature, how many of the samples are phishing, suspicious, or legitimate.
	
	For understanding the dataset better, we explain each feature briefly. For this purpose, we used references such as research of \cite{mohammad2015phishing} and a Computer Networks reference book \cite{tanenbaum2003computer}.
	
	\subsubsection*{Having$\_$IP$\_$Address}
	
	If a URL includes an IP in the domain name, the website is phishing. It should be noticed that some times the IP will be turned into a hexadecimal code, so it will be hard for users to pay attention to it easily.
	
	\subsubsection*{URL$\_$Length}
	
	Phishers use long URLs in order to hide the suspicious part in the address bar. Studies show that if the length of a URL is less than 54 characters, the website is legitimate with a high probability. If the length is more than 54 and less than 75 characters, the website is suspicious, and if the length is more than 75 characters, the website is more probable to be phishing.
	
	\subsubsection*{Shortening$\_$service}
	
	Shortening URL is a method that shortens the URL address significantly. First, by clicking on the shortened URL, users would be referred to the website which offers this service, and then enters the main website. In this dataset, a shortened URL is considered as a probable sign of a phishing website.
	
	\subsubsection*{Having$\_$At$\_$symbol}
	
	If a URL includes the symbol "@", the website is phishing, and if not, it is legitimate.
	
	\subsubsection*{Double$\_$Slash$\_$Redirecting}
	
	If a URL includes the symbol "//", it is a sign of redirecting users to a new website. Studies show that if a URL starts with HTTP, the symbol "//" should be in the sixth position, and if it starts with HTTPS, the symbol "//" should be in the seventh position. Thus, if the symbol "//" is in the next positions, the website is phishing. Otherwise, it is legitimate.
	
	\subsubsection*{Prefix-Suffix}
	
	It rarely happens that a legitimate URL includes the symbol "-". Phishers usually put a prefix or a suffix isolated by a "-" in the URL, so users assume that they are facing a legitimate website.
	
	\subsubsection*{Having$\_$Sub$\_$Domain}
	
	Consider the URL "http://www.hud.ac.uk/students". By omitting "www." from it and counting dots in the remaining part, we can bring up with a rule. If the number of remaining dots is 1, the website is legitimate, if 2, the website is suspicious (because it has subdomain) and if the number is greater than 2 (it has multi-subdomains), the website is more probable to be phishing.
	
	\subsubsection*{SSLfinal$\_$State}
	
	If a URL supports HTTPS, it significantly has improved the probability of being legitimate. However, the existence of HTTPS is not enough itself, and for more assurance, features like SSL's (Secure Sockets Layer) source and its certification age should be considered. Studies show that if a website uses HTTPS and the certificate's sources are valid, and its certificate age is more than one year, the website is legitimate, and if the website uses HTTPS but the certificate's sources are not valid, it is suspicious to be phishing. Otherwise, it is phishing with a high probability.
	
	\subsubsection*{Domain$\_$Registration$\_$Length}
	
	Most phishing websites will not be on a specified domain on the World Wide Web for a long time, while if a website is a legitimate one and wants to be on a specified domain for a long time, the cost will be prepaid. Thus, if a website wants to be on a specified domain for a short time, it is more probable that it is phishing. Otherwise, it is legitimate.
	
	\subsubsection*{Favicon}
	
	Favicon is a particular graphical image that would be placed as an icon beside the address bar. If this icon is loaded from a domain other than the website domain, the website is more probable to be phishing. Otherwise, it is legitimate.
	
	\subsubsection*{Port}
	
	When a user sends a request to a particular server, along with the request, the number of expected ports for answering back from the server will be sent. For protecting the user's information, websites must have a particular control on the ports. If all ports are open, hackers can threaten the user's information. If all ports are closed except 80 and 443, it reduces the probability of any break-in, and the website is more probable to be legitimate.
	
	\subsubsection*{HTTPS$\_$Token}
	
	Phishers will use HTTPS in the URL but not in the right place. They put it after HTTP to make the website feel legitimate. For example, we can take a look at a URL like this: 
	http://https-www-paypal-it-webapps-mpp-home.soft-hair.com/
	
	So, if there is HTTPS in the domain, the website is phishing. Otherwise, it is legitimate.
	
	\subsubsection*{Request$\_$URL}
	
	In legitimated websites that contain photos, videos, and such things, their source and the website's URL should have the same domain. In general, if the percentage of different cases is less than 22$\%$, the website is legitimate, and if it is between 22$\%$ and 61$\%$, the website is suspicious. Otherwise, the website is more probable to be phishing.
	
	\subsubsection*{URL$\_$Of$\_$Anchor}
	
	If we need to make a link from our website to another website, we use the "$<$a$>$" tag. Thus, there are two situations:
	
	\begin{enumerate}
	    \item 
    The "$<$a$>$" tag's domain is different from the website's domain
    
    \item The "$<$a$>$" tag is not linking any website (for example: $<$a href="$\#$"$>$)
    
    \end{enumerate}
    
    If the percentage of any of the explained situations is less than 31$\%$ of the whole HTML code, the website is legitimate. If this percentage is between 31$\%$ and 67$\%$, the website is suspicious. Otherwise, the website is more probable to be phishing.
	
	\subsubsection*{Links$\_$In$\_$Tags}
	
	In HTML programming language, programmers use "$<$Meta$>$", "$<$Script$>$" and "$<$Link$>$" tags in HTML documents. In legitimate websites, it is expected that these tags have their links in the same website domain. If the percentage of differences in the domains is less than 17$\%$, the website is legitimate. If it is between 17$\%$ and 81$\%$, the website is suspicious. Otherwise, it is more probable that the website is phishing.
	
	\subsubsection*{SFH (Server Form Handler)}
	
	It is a field that contains an address that the user receives from the server. If SFH is empty, the website is phishing. If its domain is different from the website domain, the website is suspicious. Otherwise, the website is legitimate.
	
	\subsubsection*{Submitting$\_$To$\_$email}
	
	If a website wants users to enter their personal information like email address  by using "mail()" or "mailto:" method, it is likely that it is just an effort to access their information. So the website is more probable to be phishing.
	
	\subsubsection*{Abnormal$\_$URL}
	
	This feature can be extracted from websites of the "WHOIS" database. If the URL does not include the host's name, the website is phishing, and if not, it is legitimate.
	
	\subsubsection*{Redirect}
	
	Legitimate websites redirect at most once. If the number of redirection is between 2 and 4, the website is suspicious, more than that, it will be phishing.
	
	\subsubsection*{On$\_$Mouseover}
	
	If the URL in status bar changes with "onMouseOver", the website is phishing. Otherwise, it is legitimate.
	
	\subsubsection*{Right$\_$Click}
	
	If Right$\_$Click is disabled on the website, it is more probable to be phishing. Otherwise, it is legitimate.
	
	\subsubsection*{Pop$\_$Up$\_$Windows}
	
	In legitimate websites, it is not common to ask users to enter their personal information in a popup window, and these windows are being used for welcoming or warning users. In general and with a high probability, if users are not asked to enter a text in pop up windows, the website is legitimate. Otherwise, it is phishing.
	
	\subsubsection*{Iframe}
	
	There is a tag in HTML that allows displaying a website on another website. Phishers may use this feature and make the frame invisible. Thus, if a website uses the $<$iframe$>$ tag, it is more probable to be phishing. Otherwise, it is legitimate.
	
	\subsubsection*{Age$\_$Of$\_$Domain}
	
	Phishing websites are usually available for a short time. Studies show that websites older than six months are legitimate. Otherwise, they are phishing. This feature can be seen on the "WHOIS" website.
	
	\subsubsection*{DNS$\_$Record (Domain Name Server Record)}
	
	This feature can be recognized from the "WHOIS" database. If this feature is empty or is not among the features in "WHOIS", the website is phishing. Otherwise, it is legitimated.
	
	\subsubsection*{Web$\_$Traffic}
	
	"Alexa" is a database that ranks websites based on their views. In the worst ranking, legitimate websites are among the top 100,000 websites. If the website is ranked less than 100,000, the website is legitimate. If it is ranked more than 100,000, it is suspicious, and if it is not among the "Alexa"'s ranking, it is more probable to be phishing.
	
	\subsubsection*{Page$\_$Rank}
	
	PageRank shows the importance of websites and gets values between 0 and 1. In the examined dataset, 95$\%$ of the phishing websites did not have PageRank, and the other 5$\%$ had a PageRank value lower than 0.2. Thus, if the PageRank value is more than 0.2, the website is legitimate. Otherwise, it is phishing.
	
	\subsubsection*{Google$\_$Index}
	
	Phishing websites are not usually in Google's index for their short availability. Thus, if a website is not on Google's index, it is more probable to be phishing. Otherwise, it is legitimate.
	
	\subsubsection*{Links$\_$Pointing$\_$To$\_$Page}
	
	In general, if there are many links from other websites to the website being tested, it is more probable that the tested website is legitimate, otherwise phishing.
	In this dataset, if there are no websites to point the website, we consider it phishing. If this number is less than 2, the website is suspicious. Otherwise, it is legitimate.
	
	\subsubsection*{Statistical$\_$Report}
	
	"Phishtank Stats" and "StopBadware" are two of the institutes working on providing statistical reports regarding phishing websites. If the website's host is in the list of phishing IPs or domains from those two institutes, the website is phishing. Otherwise, it is legitimate.
	
	\begin{figure}[htbp]
		\centering
		\includegraphics[width=0.25\textwidth]{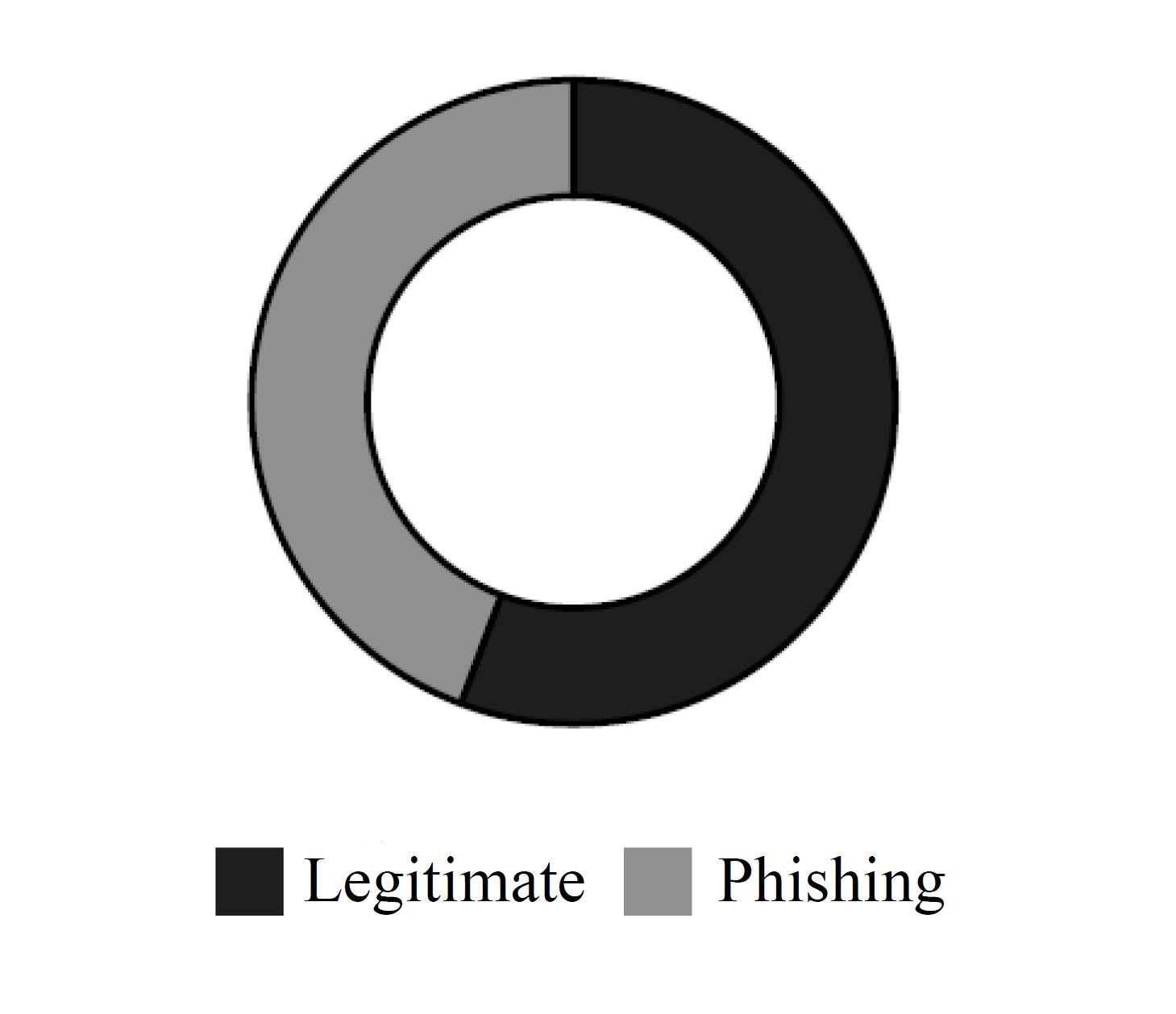}
		\caption{The proportion of phishing and legitimate websites in the dataset}
		\label{fig:classpiechart}
	\end{figure}
    
	After a brief overview of the features, we review the data labels. We show the phishing class label by -1 and legitimate class label by 1 in the dataset. The number of phishing classes is 4898, and the number of legitimate classes is 6157. In \figurename{ \ref{fig:classpiechart}} the chart that shows the proportion of phishing classes and legitimate classes in the whole dataset, is presented.

\section*{List of abbreviations}
        
    HEFS: Hybrid Ensemble Feature Selection; HTTP: Hypertext Transfer Protocol; URL: Uniform Resource Locator; HTTPS: Hypertext Transfer Protocol Secure; SSL: Secure Sockets Layer; IP: Internet Protocol; SFH: Server Form Handler; DNS: Domain Name System; HTML: HyperText Markup Language; MST: Maximum Spanning Tree

\bibliography{bibfile.bib}
\bibliographystyle{IEEEtran}

\end{document}